\documentclass[apjl]{emulateapj}
\usepackage{amsfonts,amsmath,graphicx,natbib,apjfonts,subfigure,gensymb}
\usepackage[normalem]{ulem}
\usepackage[usenames,dvipsnames]{color}
\usepackage{soul}

\def\nw{1}
\def\fer{2}
\def\msu{3}
\def\cfa{4}
\def\hbh{5}
\def\yu{6}
\def\ou{7}
\def\co{8}
\def\hubble{9}
\def\nyu{10}
\def\laim{11}
\def\pu{12}
\def\nsf{13}

\shorttitle{Jets and relativistic outflows in SLSNe-I: Radio Constraints}
\shortauthors{Coppejans et al.}

\begin{document}
\title{Jets in Hydrogen-poor Super-luminous Supernovae:\\Constraints from a comprehensive analysis of radio observations}

\author{D.~L. Coppejans\altaffilmark{\nw}, R. Margutti\altaffilmark{\nw}, C. Guidorzi\altaffilmark{\fer}, L. Chomiuk\altaffilmark{\msu}, K.~D. Alexander\altaffilmark{\cfa}, E. Berger\altaffilmark{\cfa}, M.~F. Bietenholz\altaffilmark{\hbh,\yu}, P.~K. Blanchard\altaffilmark{\cfa}, P. Challis\altaffilmark{\cfa}, R. Chornock\altaffilmark{\ou}, M. Drout\altaffilmark{\co}, W. Fong\altaffilmark{\nw,\hubble}, A. Mac Fadyen\altaffilmark{\nyu}, G. Migliori\altaffilmark{\laim}, D. Milisavljevic\altaffilmark{\pu}, M. Nicholl\altaffilmark{\cfa}, J.~T. Parrent\altaffilmark{\cfa}, G. Terreran\altaffilmark{\nw}, B.~A. Zauderer\altaffilmark{\nyu,\nsf}}

\altaffiltext{\nw}{Center for Interdisciplinary Exploration and Research in Astrophysics (CIERA) and Department of Physics and Astronomy, Northwestern University, Evanston, IL 60208}
\altaffiltext{\fer}{Department of Physics and Earth Science, University of Ferrara, via Saragat 1, I-44122, Ferrara, Italy}
\altaffiltext{\msu}{Department of Physics and Astronomy, Michigan State University, East Lansing, MI:48824, USA}
\altaffiltext{\cfa}{Harvard-Smithsonian Center for Astrophysics, 60 Garden St., Cambridge, MA 02138, USA}
\altaffiltext{\hbh}{Hartebeesthoek Radio Observatory, PO Box 443, Krugersdorp, 1740, South Africa}
\altaffiltext{\yu}{Department of Physics and Astronomy, York University, Toronto, M3J 1P3, Ontario, Canada}
\altaffiltext{\ou}{Astrophysical Institute, Department of Physics and Astronomy, 251B Clippinger Lab, Ohio University, Athens, OH 45701, USA}
\altaffiltext{\co}{Carnegie Observatories, 813 Santa Barbara Street, Pasadena, CA 91101, USA}
\altaffiltext{\hubble}{Hubble Fellow}
\altaffiltext{\nyu}{Center for Cosmology and Particle Physics, New York University, 4 Washington Place, New York, NY 10003, USA}
\altaffiltext{\laim}{Laboratoire AIM (CEA/IRFU - CNRS/INSU - Universit\'{e} Paris Diderot), CEA DSM/IRFU/DAp, F-91191 Gif-sur-Yvette, France.}
\altaffiltext{\pu}{Department of Physics \& Astronomy, Purdue University, West Lafayette, IN 47907 U.S.A.}
\altaffiltext{\nsf}{National Science Foundation, 2415 Eisenhower Avenue Alexandria, VA 22314}

\begin{abstract}
The energy source powering the extreme optical luminosity of hydrogen-stripped Superluminous Supernovae (SLSNe-I) is not known, but recent studies have highlighted the case for a central engine. Radio and/or X-ray observations are best placed to track the fastest ejecta and probe the presence of outflows from a central engine. We compile all the published radio observations of SLSNe-I to date and present three new observations of two new SLSNe-I. None were detected. Through modeling the radio emission, we constrain the sub-parsec environments and possible outflows in SLSNe-I. In this sample we rule out on-axis collimated relativistic jets of the kind detected in Gamma-Ray Bursts (GRBs). We constrain off-axis jets with opening angles of 5\arcdeg\ (30\arcdeg) to energies of $\rm{E_k<4\times10^{50}\,erg}$ ($\rm{E_k<10^{50}\,erg}$) in environments shaped by progenitors with mass-loss rates of $\dot{M}<10^{-4}\,M_{\odot}\,{\rm yr}^{-1}$ ($\dot{M}<10^{-5}\,M_{\odot}\,{\rm yr}^{-1}$) for all off-axis angles, assuming fiducial values $\epsilon_e=0.1$ and $\epsilon_B=0.01$. The deepest limits rule out emission of the kind seen in faint un-collimated GRBs (with the exception of GRB\,060218), and from relativistic supernovae. Finally, for the closest SLSN-I SN 2017egm we constrained the energy of an uncollimated non-relativistic outflow like those observed in normal SNe to $E_{\rm k}\lesssim10^{48}$ erg.

\end{abstract}



\section{Introduction}
\setcounter{footnote}{13}

Super Luminous Supernovae (SLSNe) are a distinct class of Supernovae (SNe) that have UV-optical luminosities $L>7\times 10^{43}\,{\rm erg \,s^{-1}}$ \citep{Quimby11,Chomiuk11}. These stellar explosions are typically $\sim$10-100 times more luminous than ordinary SNe\footnote{Note that \citet{Milisavljevic13} and \citet{Lunnan17} found SNe with spectroscopic similarities to SLSNe, but at lower luminosities.}, show comparatively bright UV emission at early times, and in some cases have decay rates that are incompatible with $^{56}$Ni and $^{56}$Co decay \citep{Gal-Yam12,Lunnan17,DeCia17}. 

There are two main classes of SLSNe, namely the Hydrogen-rich systems (SLSNe-II) and the Hydrogen-stripped systems (SLSNe-I). Some SLSNe-II show clear signatures of shock interaction with a dense medium in their optical spectra (in the form of narrow emission lines with width $\rm{<100~km\,s^{-1}}$). For these systems, the large UV-optical luminosity can be explained through the interaction of the blast wave with dense material left behind by the stellar progenitor before collapse \citep[e.g.,][]{Smith07,Ofek07,Chatzopoulos11}. The mechanism, or mechanisms, that power the exceptional luminosities of Hydrogen-stripped systems (SLSNe-I) however, are unknown \citep[e.g.,][]{Gal-Yam12}.

A number of models for the energy source of SLSNe-I have been proposed. Higher luminosities could be explained by the presence of larger quantities of radioactive material (with respect to ordinary SNe), or else a central engine. Large quantities of $^{56}$Ni could be produced by a pair instability supernova \citep{Woosley07,Gal-Yam09}. A central engine in the form of the spin-down of a magnetar \citep[e.g.,][]{Kasen10,Woosley10,Nicholl13,Metzger15}, or fallback accretion onto the compact remnant \citep{Dexter13} has been suggested. The source of the large luminosity could also be due to increased efficiency of the conversion of kinetic energy into radiation via shock interaction in a particularly dense circumstellar medium \citep[e.g.,][]{Smith07, Chevalier11, Ginzburg12}. 

The mechanism (or mechanisms) powering the luminosity of SLSNe-I are a topic of debate. A key problem with the interaction model is that no clear evidence for a dense surrounding medium (such as narrow spectral lines with $v\leq100$ km\,s$^{-1}$ at early times) has been observed in SLSNe-I. \citet{Roth16} however, show that under the right conditions the narrow line emission could be suppressed. H-alpha emission has been detected at late times in three SLSNe-I \citep{Yan15,Yan17b}: to power this emission and not produce narrow lines, a few solar masses of hydrogen-free material would need to have been ejected in the last $\sim$year before stellar explosion \citep[e.g.,][]{Chevalier11,Ginzburg12,Chatzopoulos12,Moriya13}. There are however claims that interaction of the ejecta with the medium is necessary to fit the light curves of some SLSNe-I, regardless of whether interaction is the dominant contribution to the flux \citep[e.g.,][]{Yan15,Wang16,Tolstov17}.

Pair Instability SNe \citep{Barkat67} could produce the required amounts of $^{56}$Ni to power the optical luminosity, but to date only two candidates are known \citep{Gal-Yam09,Terreran17} and the classification is debated in the literature \citep[e.g.,][]{Yoshida11,Nicholl13}. If the radioactive decay of $^{56}$Ni is the sole energy source, then for some SLSNe-I, the necessary quantities cannot be reconciled with the inferred ejecta mass, bright UV emission or the decay rate of the light curves (e.g., \citealt{Kasen11,Dessart12,Inserra13,Nicholl13,McCrum14}; see however \citealt{Kozyreva17}). Pair Instability explosions cannot account for the entire class of SLSN-I.

Recent studies are increasingly favoring the central engine model \citep[e.g.,][]{Nicholl17c,Margalit17} as it has been shown to satisfactorily reproduce the optical light-curves of SLSNe-I with a wide range of properties \cite[e.g.,][]{Inserra13,Chatzopoulos13,Nicholl14,Metzger15,Inserra17,Nicholl17}. Magnetar central engines with initial spin periods in the range 1--5 ms and magnetic fields in the range $\approx 10^{13}-10^{14}$ G are the best fit for the optical bolometric emission of several systems \citep[e.g.,][]{Dessart12,Inserra13,Nicholl13,Metzger15,Lunnan16,Yu17}. 

\begin{deluxetable*}{lccccll}
\tabletypesize{\scriptsize}
\tablecolumns{7} 
\tablewidth{40pc}
\tablecaption{Properties of the sample of SLSNe-I}
\tablehead{  \colhead{Name(s)} & \colhead{$d_{L}$}& Explosion Date & \colhead{Time after expl.$^{*}$}& \colhead{Freq.$^{*}$}& \colhead{Specific Luminosity}& \colhead{Reference}\\
& (Mpc) & (MJD) & (days) & (GHz) & (erg/s/Hz) &}
\startdata
PTF09cnd & 1306$^{a}$ & 55006$^{b}$ & 85 & 10.6 & <1.5$\times 10^{29}$ & \citealt{Chandra09ATel}\\
 &  &  & 85 & 6.1 & <1.6$\times 10^{29}$ & \citealt{Chandra09ATel}\\
 &  &  & 85 & 1.8 & <9.4$\times 10^{31}$ $^{**}$ & \citealt{Chandra09ATel}\\
 &  &  & 140 & 10.6 & <1.1$\times 10^{29}$ & \citealt{Chandra10ATel}\\
 & &  & 142 & 6.1 & <2.4$\times 10^{29}$ & \citealt{Chandra10ATel}\\
 &  &  & 147 & 1.8 & <9.7$\times 10^{29}$  $^{**}$ & \citealt{Chandra10ATel}\\
PS1-10awh & 5865 & 55467 & 39 & 9.6 & <9.7$\times 10^{29}$ & \citealt{Chomiuk11}\\
PS1-10bzj & 3891$^{c}$ & 55523$^{c}$ & 48 & 8.2 & <8.6$\times 10^{29}$ & This work\\
PS1-10ky & 6265 & 55299$^{d}$ & 156 & 9.6 & <1.2$\times 10^{30}$ & \citealt{Chomiuk11}\\
SN 2012il$^e$ & 825$^{f}$ & 55919$^g$ & 44 & 6.9 & <1.5$\times 10^{28}$ & \citealt{Chomiuk12ATel}\\
SN 2015bn$^h$ & 528 & 57013$^{i}$ & 318 & 8.2 & <2.2$\times 10^{28}$ & \citealt{Nicholl16}\\
 &  &  & 318 & 24.5 & <1.2$\times 10^{28}$  $^{**}$ & \citealt{Nicholl16}\\
iPTF15cyk & 3101$^{j}$ & 57249$^{k}$ & 61 & 8.3 & <2.2$\times 10^{29}$ & \citealt{Palliyaguru16,Kasliwal16}\\
 &  &  & 94 & 8.3 & <1.7$\times 10^{29}$ & \citealt{Palliyaguru16,Kasliwal16}\\
 &  &  & 124 & 8.3 & <1.7$\times 10^{29}$ & \citealt{Palliyaguru16,Kasliwal16}\\
Gaia16apd$^l$ & 467$^{m}$ & 57512$^{n}$ & 26 & 6.6 & <4.7$\times 10^{27}$ & This work\\
 &  &  & 26 & 24.0 & <1.1$\times 10^{28}$  $^{**}$ & This work\\
 &  &  & 203 & 6.6 & <3.6$\times 10^{27}$ & This work\\
 &  &  & 203 & 24.0 & <7.6$\times 10^{27}$  $^{**}$ & This work\\
SN 2017egm$^o$ & 136$^{p}$ & 57887$^{q}$ & 34 & 16.0 & <3.9$\times 10^{28}$  $^{r,**}$ & \citealt{Bright17ATel} \\
 &  &  & 38 & 5.2 & <1.3$\times 10^{27}$  $^{**}$ & \citealt{Bright17ATel} \\
 &  &  & 39 & 10.3 & <5.7$\times 10^{26,s}$ & \citealt{Romero-Canizales17ATel,Bose17}\\
 &  &  & 39 & 1.6 & <2.1$\times 10^{27}$  $^{**}$ & \citealt{Romero-Canizales17ATel,Bose17}\\
 &  &  & 46 & 34.0 & <3.3$\times 10^{27}$  $^{**}$ &  This work; \citealt{Coppejans17ATel}\\
  &  &  & 47 & 10.3 & <6.4$\times 10^{26,s}$ & \citealt{Romero-Canizales17ATel,Bose17}\\
  &  & & 74 & 34.0 & <7.4$\times 10^{26}$  $^{**}$ &  This work\\
 \hline
\multicolumn{7}{p{40pc}}{Notes: 3$\sigma$ Upper-limits are given on the luminosity. $^{*}$Explosion rest frame. $^{**}$Not used in the modelling of emission from off-axis jets (Section \ref{sec:jets}), because the explosion rest frame frequency is significantly different from $\sim8$ GHz. $^{a}$\citet{Neill11}. $^{b}$From \citet{Quimby11} the peak time is MJD 55069.145, and assuming a rest-frame rise-time of 50 days. $^{c}$The peak was at MJD 55563.65+-2 \citep{Lunnan13}, and as this was a fast rising SN (Lunnan et al. in prep), we assume a rest-frame rise time of 25 days. $^{d}$Calculated based on the peak time from \citet{Chomiuk11}, and assuming a 50-day rise time in the explosion rest frame. $^e$Aliases: PS1-12fo and CSS120121:094613+195028. $^{f}$\cite{Smartt12ATel,Inserra13}. $^{g}$\citet{Inserra13}. $^h$Aliases: PS15ae, CSS141223:113342+004332 and MLS150211:113342+004333. $^{i}$The SN reached $r$ band maximum light on MJD 57102, and the inferred rise-time in the explosion rest-frame is $\sim80$ days \cite{Nicholl16}. $^{j}$\cite{Kasliwal16}. $^{k}$We estimated the peak time at 57293.5 MJD based on a comparison to LSQ12dlf (private communication with Alessandra Corsi), and assumed a rise-time of 50 days in the explosion rest frame. $^l$Alias: SN 2016eay. $^{m}$\cite{Nicholl17}. $^{n}$Time of maximum light was MJD 57541 \cite{Yan15}, and rise-time in the rest-frame was 29 days \cite{Nicholl17}. $^o$Alias: Gaia17biu. $^{p}$\citet{Romero-Canizales17ATel}. $^{q}$\citet{Nicholl17b}. $^{r}$As the contribution of the galaxy (NGC 3191) to the detected flux density is unknown, we take it as an upper-limit on the SN. $^{s}$These observations are presented in \citet{Bose17} and \citet{Romero-Canizales17ATel}. Individual upper-limits for the two observations are from private communication with Subhash Bose and Cristina Romero-Canizales.}
\enddata
\label{Tab:Sample}
\end{deluxetable*}

\begin{deluxetable*}{lcccccccc}
\tabletypesize{\scriptsize}
\tablecolumns{9} 
\tablewidth{40pc}
\tablecaption{Details of Observing Runs}
\tablehead{  \colhead{Name(s)} & \colhead{Start Obs.} & \colhead{Total Obs.$^a$} & \colhead{Frequency} & \colhead{Bandwidth} & \colhead{VLA Array} & \colhead{Beam size} & \colhead{Beam Angle} & \colhead{Flux Density}\\
 & (MJD) & Time (s) & (GHz) & (GHz) & Config. & FWHM (arcsec) & (deg) & ($\mu$Jy)}
\startdata
PS1-10bzj & 55603.08146 & 2483 & 4.96 & 0.256 & C-CnB & 2.88$\times$1.30 & $-167.8$ & <87\\
Gaia16apd$^c$ & 57541.32332 & 1730 & 5.9 & 2.048 & B & 2.28$\times$1.07 & 69.1 & <20.4\\
& 57541.30115 & 1731 & 21.8 & 2.048 & B & 0.57$\times$0.31 & 74.5 & <45.9\\
& 57736.49994 & 1730 & 5.9 & 2.048 & A & 0.35$\times$0.32 & 41.9 & <15.3\\
& 57736.47778 & 1731 & 21.8 & 2.048 & A & 0.11$\times$0.09 & 58.7 & <32.1\\
SN 2017egm$^{b}$ & 57933.96987 & 2052 & 33.0 & 8.192 & C & 0.76$\times$0.62 & 16.6 & <150$^{d,e}$\\
& 57962.69774 & 2052 & 33.0 & 8.192 & C & 0.97$\times$0.64 & -87.4 & <33.6\\
\hline
\multicolumn{9}{p{40pc}}{Notes: Upper-limits are 3$\sigma$. $^a$Including intervening calibrator scans, but excluding initial setup scans. $^b$Alias: Gaia17biu. $^c$Alias: SN\,2016eay. $^d$This observation was reported in an Astronomer's Telegram \citep{Coppejans17ATel}. $^e$This observation had high noise levels due to poor weather conditions.}
\enddata
\label{Tab:Observinglog}
\end{deluxetable*}

There is growing evidence of a link between jetted long Gamma-Ray Burst (GRB) SNe and SLSNe-I in the form of observational similarities in their spectra and light curves \citep[e.g.,][]{Greiner15,Metzger15,Kann16,Nicholl16,Jerkstrand17}, their preference for metal-poor host galaxies (e.g., \citealt{Lunnan14,Perley16,Chen15,Chen17,Chen17b,Izzo17}, but also see \citealt{Chen17c,Nicholl17b,Bose17})\footnote{See \citet{Lunnan14,Leloudas15b,Angus16,Schulze2018} for a comparison of host properties.}, and in the models for the central engine \citep[e.g.,][]{Metzger15,Margalit17}. SLSNe and GRB-SNe have broader spectral features than normal H-stripped SNe indicative of large photospheric velocities \citep{Liu17b}. Additionally, the luminous UV emission in the SLSN-I Gaia16apd has been suggested to originate from a central engine (\citealt{Nicholl17}, see however \citealt{Yan17}). In the SLSN-I SCP06F6, luminous X-ray emission (outshining even GRBs at a similar post-explosion time by a large factor), \emph{if} indeed associated with the transient, is likely powered by a central engine \citep{Gansicke09,Levan13,Metzger15}. The presence of a central engine may also provide an additional driving force for the stellar explosion \citep[e.g.,][]{Soker17,Lui17}.

A key manifestation of a central engine is an associated jet. The search for evidence of a jet is best conducted at radio and X-ray wavelengths. Optical emission is of thermal origin and tracks the slowly moving material in the explosion ($v\leq$ a few $10^{4}\,\rm{km\,s^{-1}}$). In contrast, radio and X-ray emission are of non-thermal origin, and arise from the interaction of the explosion's fastest ejecta ($v\geq 0.1c$)   with the local environment. As the radiative properties of the shock front are directly dependent on the circumstellar density, radio/X-ray observations also probe the mass-loss history of the progenitor star in the years prior to explosion, a phase in stellar evolution that is poorly understood (see \citealt{Smith14} for a recent review).

In \citet{Margutti17b} we used the sample of X-ray observations of SLSNe-I to constrain relativistic hydrodynamical jet models and determine constraints on the central engines and sub-parsec environments of SLSNe-I. This work showed that interaction with a dense circumstellar medium is not likely to play a key role in powering SLSNe-I, and that at least some SLSNe-I progenitors are compact stars surrounded by a low density environment. There was no compelling evidence for relativistic outflows, but the limits were not sensitive enough to probe jets that were pointed more than 30$\degr$ out of our line of sight. In one case (PTF12dam) the X-ray limits were sufficiently deep to rule out emission similar to sub-energetic GRBs, suggesting a similarity to the relativistic SNe 2009bb and 2012ap \citep{Soderberg10,Chakraborti15,Margutti14b} if this SLSN-I was a jet-driven explosion. 

In this paper, we expand on recent analysis of radio observations from the SLSN\,2015bn \citep{Nicholl16,Margalit17} and compile all the radio-observed SLSNe-I including the three new observations presented for the first time in this work, with the aim of placing stronger constraints on the properties of their sub-pc environment and fastest ejecta (both in the form of a relativistic jet and an uncollimated outflow). These data span $\sim26-318$ days after the explosion (in the explosion rest frame, and at GHz frequencies). We test for the presence of a central engine that would produce a GRB-like jet and model the on-axis and off-axis emission from a jet for a range of densities, micro-physical shock parameters, kinetic energies, jet opening angles, and off-axis observer angles and compare these to observations. We also explore the radio properties of uncollimated outflows that would be consistent with our limits and derive constraints on the fastest ejecta and mass-loss history of SLSNe-I.

In section \ref{sec:sample} we describe the sample of radio-observed SLSNe-I. In section \ref{sec:observations} we present our new radio observations of SLSNe-I and provide details on the data reduction. The constraints on on-axis and off-axis jets are given in \ref{sec:jets}. Section \ref{sec:uncollimated_outflows} describes the constraints we derive for uncollimated outflows. Conclusions are drawn in Section \ref{sec:Conclusions}.

Unless otherwise stated, all time intervals and frequencies are quoted in the explosion rest frame and the error bars are 1-sigma. We assume a Lambda cold dark matter cosmology with $H_0=70$ km\,s$^{-1}$\,Mpc$^{-1}$ ($h=0.7$), $\Omega_0=0.3$, $\Omega_{\Lambda}=0.7$. 

\section{Sample}\label{sec:sample}

Our sample consists of all H-stripped SLSNe with published radio observations as of August 2017, comprising nine SLSNe-I. This includes seven systems with radio observations already published in the literature (PS1-10ky, PS1-10awh, PS1-12fo, iPTF15cyk, SN 2015bn, PTF09cnd and SN 2017egm) and two systems (PS1-10bzj, and Gaia16apd) for which we present the first radio observations. We also present the latest observations of SN 2017egm, updating the observations from \citet{Bose17}. A brief description of each SLSN is given in Appendix \ref{appendix:descriptions}. Table \ref{Tab:Sample} gives the radio observations and relevant references for all these systems.

The exact date of explosion is not known for every object in this sample. In a number of cases, the time of the observations (in number of days since the explosion) were derived from the time of peak luminosity and an estimated rise-time. Given the spread in rise-times for SLSNe-I \citep[see][]{Nicholl17c}, the uncertainty on these derived observation times is less than 20 days. We tested the impact of this uncertainty on our conclusions in the following analysis by increasing, and then decreasing, the assumed rise-times of all objects in our sample by 20 days. The difference in the constraints that we derive are marginal and do not affect our conclusions.

\section{Observations}\label{sec:observations}

Our observations of SN 2017egm (NRAO observing code VLA/17A-466, PI: R. Margutti), Gaia16apd (VLA/16A-476, PI: R. Margutti) and PS1-10bzj (VLA/AS1020, PI: A. Soderberg) were taken with the Karl G. Jansky Very Large Array (VLA, \citealt{Perley11}). Table \ref{Tab:Observinglog} shows the
details of the observations. These data were calibrated using the integrated VLA pipeline in \textsc{CASA}\footnote{Common Astronomy Software Applications package \citep{McMullin07}.} v4.7.0. The observations were taken in standard phase referencing mode, and the absolute flux density scale was set via observations of a standard flux density calibrator (3C286 for Gaia16apd and 3C147 for PS1-10bzj and SN 2017egm) using the coefficients of \citet{Perley-Butler2013} which are within \textsc{CASA}. We used Briggs weighting with a robust parameter of 1 to image. Two Taylor terms were used to model the frequency dependence of the larger bandwidth observations (SN 2017egm and Gaia16apd). None of the sources were detected. We quote upper-limits as 3 times the noise level in the vicinity of the source as derived from the \textsc{CASA} Imfit task. Our results are summarized in Table \ref{Tab:Sample}, Table \ref{Tab:Observinglog} and Fig. \ref{Fig:SLSNeGRBSNe}.

\section{Constraints on relativistic jets}\label{sec:jets}

\subsection{On-axis relativistic jets}\label{sec:on-axis_jets}

\begin{figure*}
	\centering
	\includegraphics[width=0.85\textwidth]{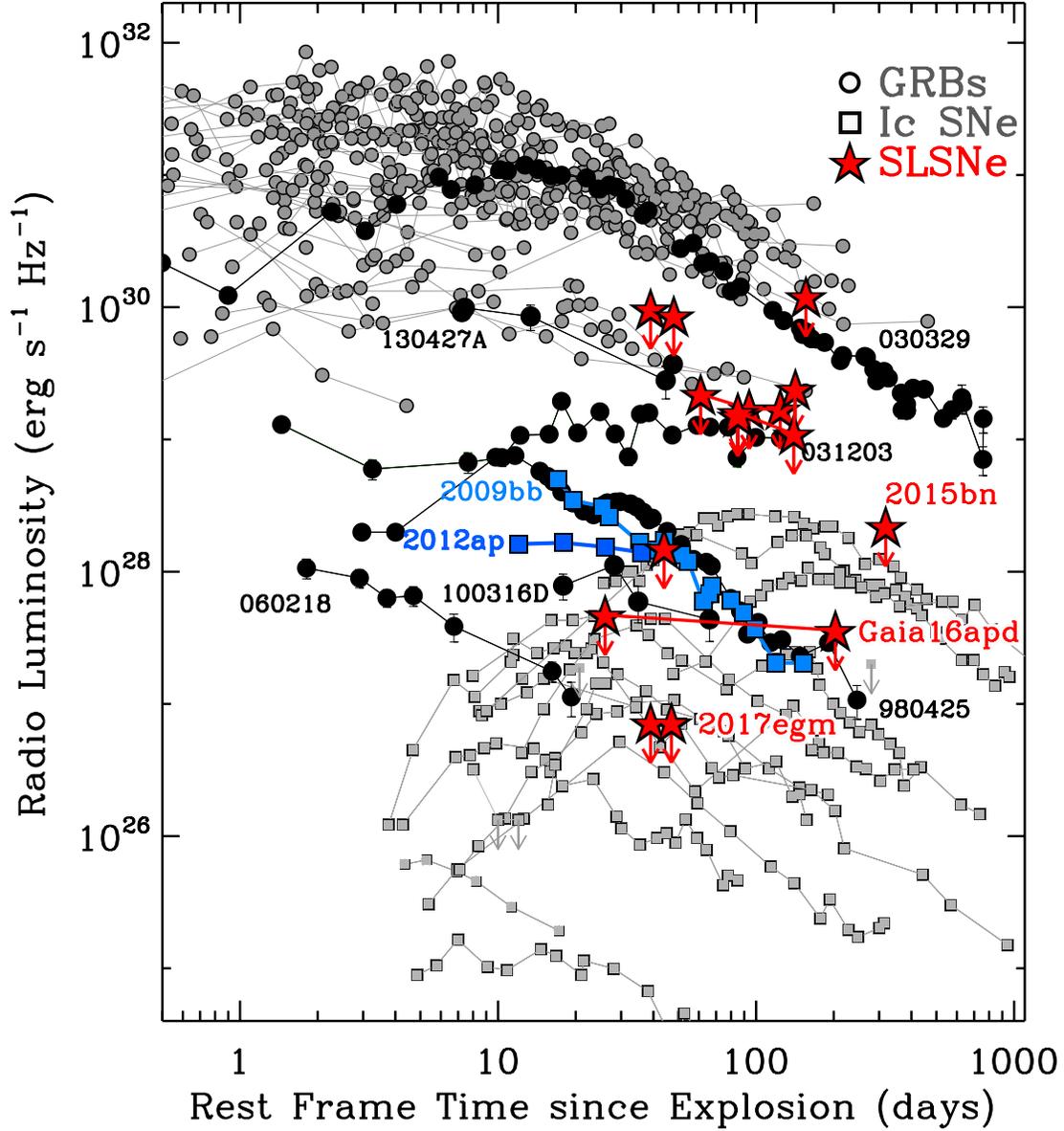}
    \caption{Specific radio luminosity at $\sim8$ GHz (rest-frame) for SLSNe-I (red stars) in the context of H-stripped core-collapse explosions (i.e. GRBs -circles- and normal Ic SNe -squares-). Black circles: GRBs at $z\leq0.3$. Grey circles: GRBs at $z>0.3$. Grey squares: normal Ic-SNe. Blue squares: relativistic Ic-SNe. Connected symbols refer to observations of the same object. For display purposes, only the SLSNe-I directly referred to in the text are labeled. Deep radio observations of the closest SLSN-I like Gaia16apd and SN 2017egm clearly rule out on-axis jets of the kind detected in GRBs, and probe the parameter space of the weakest engine driven SNe (like those associated with GRBs 980425 and 100316D). Notably, radio observations of Gaia16apd and SN 2017egm indicate that SLSNe-I can be significantly fainter than normal H-stripped core-collapse SNe as well. References: \citet{Immler02,Pooley04,Soria04,Soderberg05,Perna08,Chandra09ATel,Chandra10ATel,Soderberg10,Corsi11,Chomiuk11,Chomiuk12ATel,Chandra12,Horesh13,Margutti13,Margutti13b,Margutti14b,Corsi14,Nicholl16,Palliyaguru16,Kasliwal16,Bright17ATel,Romero-Canizales17ATel,Bose17,Coppejans17ATel}.}
\label{Fig:SLSNeGRBSNe}
\end{figure*}

Figure \ref{Fig:SLSNeGRBSNe} shows the $\sim6-10$ GHz SLSNe-I radio luminosity upper-limits in reference to those from other classes of massive stellar explosions from H-stripped progenitors, including Long GRBs (hereafter referred to just as GRBs), `normal' H-poor core-collapse SNe (Type Ibc, see \citealt{Filippenko97}), and relativistic SNe. On-axis jets in GRBs (both the collimated, and poorly collimated systems like sub-energetic GRBs 980425, 060218 and 100316D in Fig.\ \ref{Fig:SLSNeGRBSNe}) produce luminous radio emission (e.g., \citealt{Kulkarni98,Chandra09}, \citealt{Soderberg06b,Soderberg10b}). 

The SLSNe-I radio luminosity limits are significantly fainter than most cosmological GRBs detected in the radio, which typically show $L_\nu\ge 10^{29}\,\rm{erg\,s^{-1}Hz^{-1}}$ (Fig. \ref{Fig:SLSNeGRBSNe}). Our deepest luminosity limits acquired for SN 2017egm are deeper than the deepest limits for the sample of radio-observed GRBs in \citet{Chandra12} (see their Figure 6). SN 2017egm (and most of our sample of SLSNe-I) is significantly closer than cosmological GRBs (see \citealt{Chandra12}). We restrict the comparison of our SLSNe radio limits to the sample of GRBs in the local Universe ($z<0.3$) which are more representative of the true demographics (at higher redshifts $z>0.3$ we are sensitive only the high energy tail of the GRB distribution)\footnote{Since the VLA upgrade, more sensitive observations of GRBs at $z<0.3$ have consistently yielded detections \citep{Zauderer12GCN, vanderhorst13GCN, Kamble15GCN, Horesh15, Laskar16GCN}.}. At $z<0.3$ we are consequently sensitive to the entire demographics of long GRBs. For Gaia16apd and SN2017egm, our limits rule out radio emission of the kind detected from GRBs in the local Universe (black points in Fig. \ref{Fig:SLSNeGRBSNe}), with the exception of the faint GRB060218 (for which there is no evidence for collimation of the fastest ejecta).

Notably, for Gaia16apd and SN\,2017egm  we can rule out emission of the kind detected from the low-luminosity GRB\,980425 associated with SN\,1998bw (for which there is also no evidence for collimation of the fastest ejecta). This is of particular relevance as \citet{Nicholl16b} found clear similarities in the nebular spectra of the SLSN 2015bn and the GRB-SN\,1998bw, which suggests a similar core structure of their stellar progenitors at the time of collapse and possibly also a similar explosion central engine. Radio observations show that this similarity does not extend to the properties of the fastest ejecta of Gaia16apd and SN\,2017egm. (We note that for SLSN\,2015bn radio observations were acquired at a much later epoch and do not constrain GRB980425-like radio emission as shown in Fig.\ \ref{Fig:SLSNeGRBSNe}, \citealt{Nicholl16}). If the deepest limits (SN\,2017egm and Gaia16apd) are excluded, the rest of the sample still rule out emission of the kind seen in some of the low-luminosity GRBs.

Radio observations of SN\,2017egm were acquired at later times than those of the faint GRBs, leaving the possibility of a GRB\,060218-like outflow in SLSNe-I still open. To determine the presence of GRB\,060218-like emission in SLSNe-I, radio observations at $\lesssim10$ days after the explosion are necessary. This can be seen by considering GRB\,060218 in Figure \ref{Fig:SLSNeGRBSNe}: in GRB\,060218 the radio luminosity declined by approximately an order of magnitude in the first $\sim30$ days after explosion, which is the earliest phase for which we have SLSNe-I radio observations. 

We conclude that this sample of SLSNe-I are not consistent with having on-axis jets of the kind detected in GRBs. The deepest SLSNe-I limits also rule out emission from weak, poorly-collimated GRBs, with the notable exception of the fast-fading GRB\,060218.

\subsection{Off-axis relativistic jets}\label{sec:off-axis_jets}

\subsubsection{Simulation Setup}

\begin{figure} 
	\centering
	\includegraphics[width=0.5\textwidth]{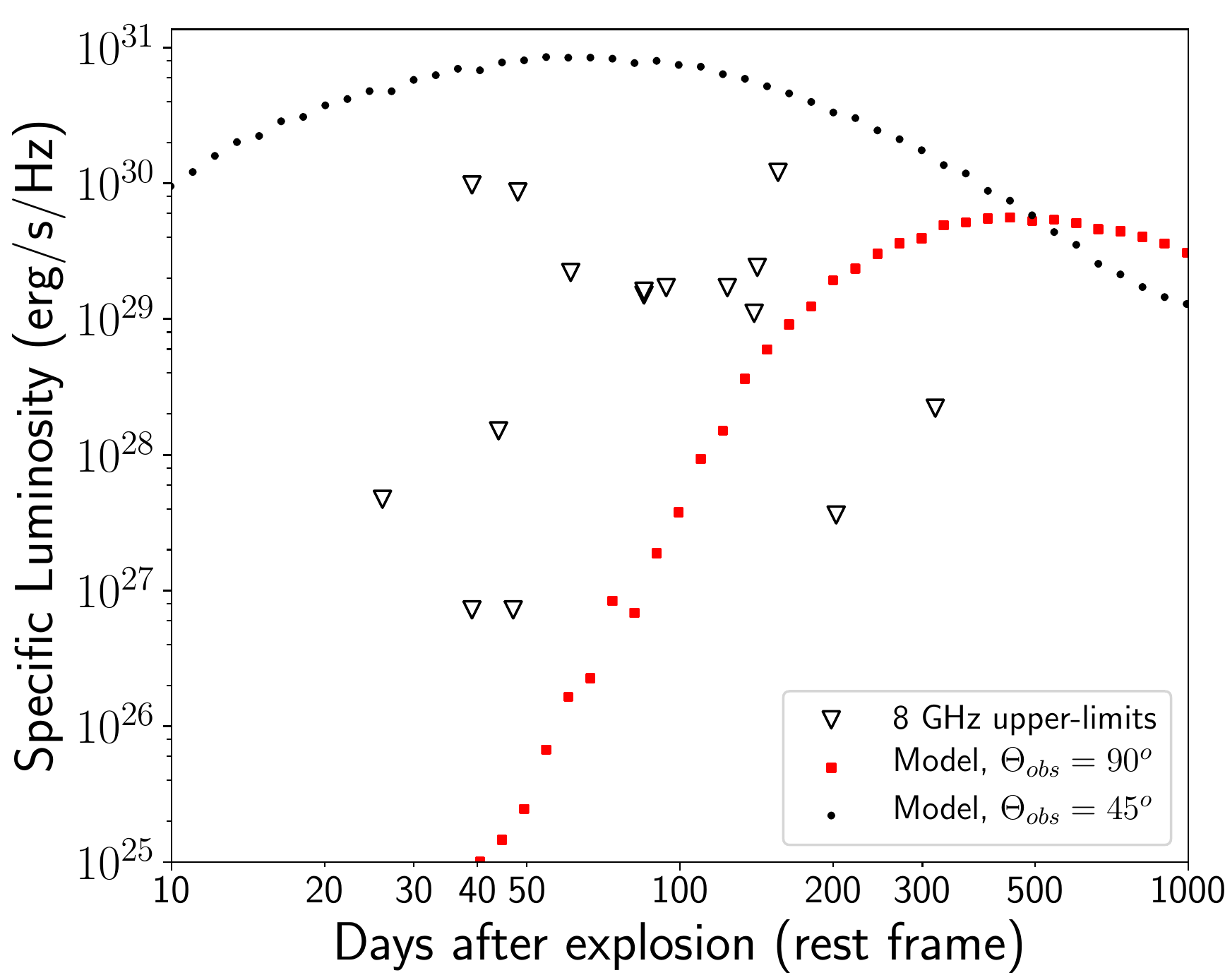}
    \caption{Example illustrating how the collective SLSNe-I $\sim$8~GHz limits and the model jet light curves are used to test a set of parameters. Two models for off-axis jets are shown for an ISM profile CSM (constant density, $\rho_{\rm CSM}$), with $\theta_{\rm j}=30\arcdeg$, $E_{\rm k,iso}=10^{53}\,{\rm erg}$, $n_{\rm CSM}=10$ cm$^{-3}$, $\epsilon_e=0.1$ and $\epsilon_B=0.01$. The red squares and black dots show the emission for the jet positioned at angles of $\theta_{\rm obs}=90\arcdeg$ and $\theta_{obs}=45\arcdeg$ respectively. The radio limits rule out this set of parameters for both models, as they are lower than the predicted specific luminosities for either angle. The radio emission from a jet at a larger off-axis angle will peak later than it would at smaller angles, as the radiation is initially beamed away from the observer and will take longer to spread into the line of sight. Late-time observations are consequently necessary to constrain off-axis jets. For this set of parameters, the latest two observations (at 203 and 318 days which are for Gaia 16apd and SN 2015bn, respectively) are more constraining for the $\theta_{obs}=90\arcdeg$ jet than the deepest radio limits, as the emission peaks later on.}
    \label{Fig:example}
\end{figure}

\begin{figure*}
	\centering
	\includegraphics[width=0.75\textwidth]{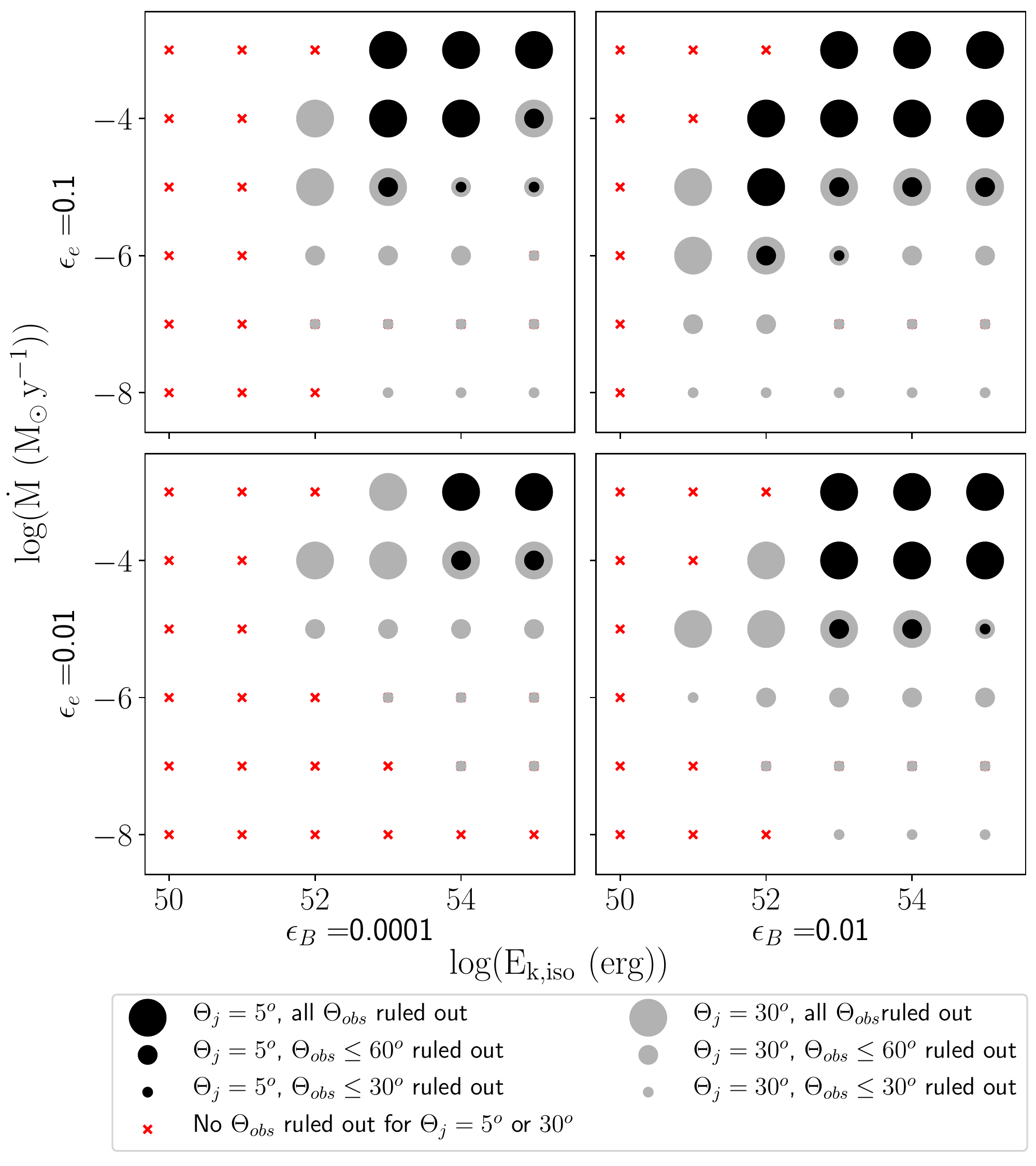}
    \caption{Constraints on jetted outflows in the sample of radio observed SLSNe-I assuming the progenitor produced a wind density profile ($\rho\propto r^{-2}$) in the surrounding medium. The symbol colours represent jet opening angles of $\theta_{\rm j}=5\arcdeg$ (black) and $\theta_{\rm j}=30\arcdeg$ (gray). Symbol sizes indicate the observer angle ($\theta_{obs}$) for which we can rule out the corresponding jet, with larger symbols corresponding to larger $\theta_{\rm obs}$. Red crosses indicate that the parameters could not be ruled out. The top (bottom) panels are $\epsilon_e=0.1$ ($\epsilon_e=0.01$), and the left (right) panels are $\epsilon_B=0.0001$ ($\epsilon_B=0.01$). Note: In the top-left panel, highly collimated jets ($\theta_{\rm j}=5\arcdeg$) with $E_{\rm k,iso} \geq 10^{53}$ erg and progenitor mass loss rates of $\dot{M} \geq 10^{-4}\,M_{\odot}\, \rm{yr}^{-1}$ are ruled out for all observer angles. The `outlier' at $E_{\rm k,iso}=10^{55}$ erg was a sampling effect where the upper-limit was negligibly more luminous than the model at $\theta_{\rm obs}=90\arcdeg$.}
    \label{Fig:windconstraints}
\end{figure*}

\begin{figure*}
	\centering
	\includegraphics[width=0.75\textwidth]{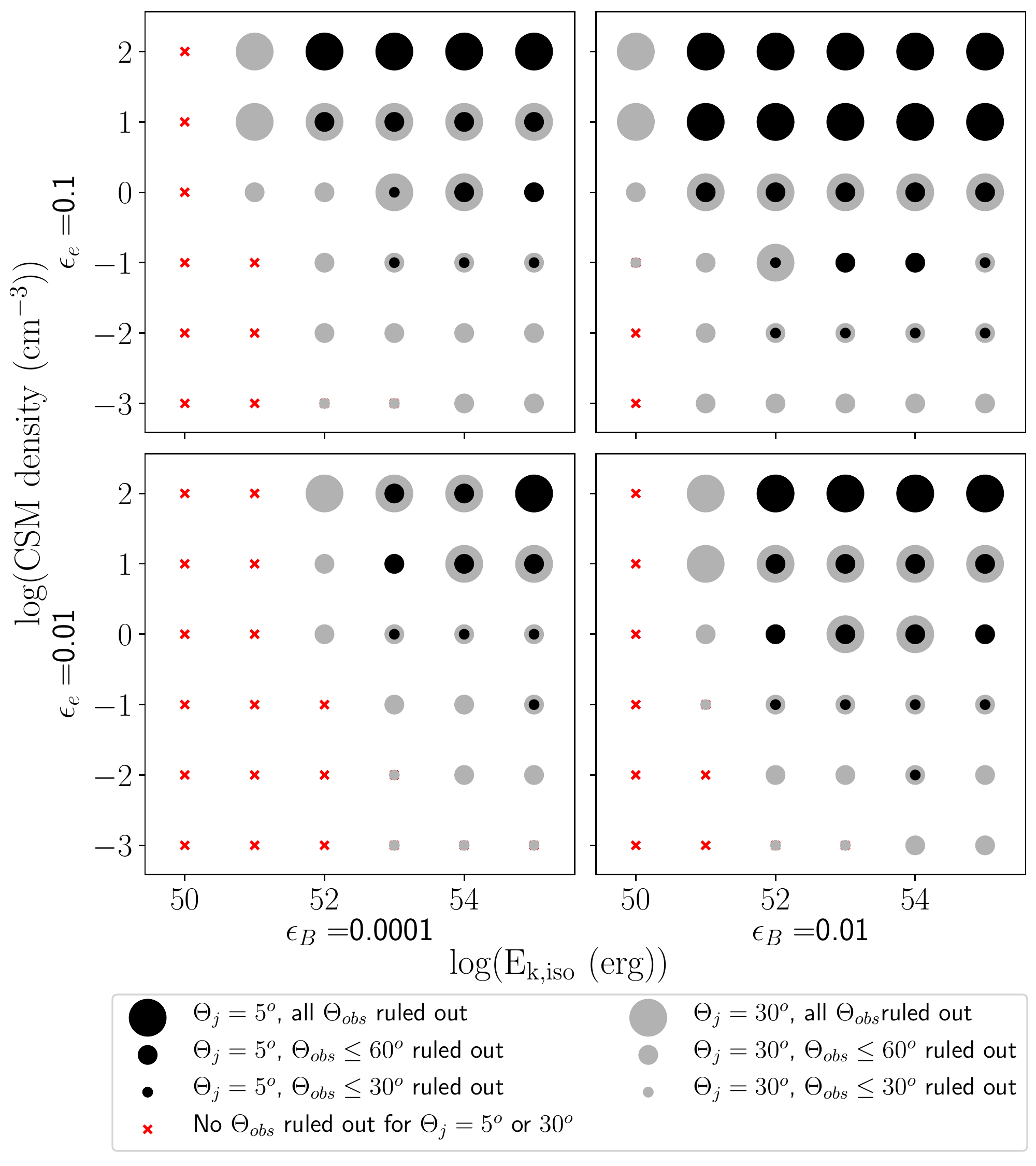}
    \caption{Constraints on jetted outflows in the sample of radio observed SLSNe-I for a constant density profile in the surrounding medium. See the caption of Figure \ref{Fig:windconstraints} for a full description of the symbols. Note: In the top-right panel, highly collimated jets ($\theta_{\rm j}=5\deg$) with $E_{\rm k,iso} \geq 10^{51}$ erg in environments with $n_{\rm CSM}=10$ cm$^{-3}$ are ruled out for all observer angles.\vspace{4mm}}
    \label{Fig:ismconstraints}
\end{figure*}

To constrain the presence of off-axis relativistic outflows in SLSNe-I, we generated a grid of model light curves for off-axis Gamma-ray Burst jets using high-resolution two-dimensional relativistic hydrodynamical jet simulations. For this, we used the broadband afterglow numerical code Boxfit v2 \citep{vanEerten12}, which models the off-axis, frequency-dependent emission as the jet slows, and the radiation becomes less beamed. We then compared the collective $\sim8$ GHz (central frequencies in the range 6.1-10.6 GHz, as indicated in Table \ref{Tab:Sample}) radio upper-limits in our sample to each light curve to determine if the observations rule out that particular set of parameters. These frequencies provide the most stringent constraints on the jet parameters, as they include the deepest limits, were taken at the earliest and latest times, and have the densest time coverage. To do this, we made the necessary assumption that every SLSN-I in our sample is powered by the same mechanism (and jet/environment properties), ie. a given set of parameters is ruled out if they are ruled out for at least one SLSN-I. The radio light curves are not sufficiently well-sampled to do this analysis individually. An illustration of this process is given in Figure \ref{Fig:example}.

The modeled radio light curves depend on the following input parameters: (1) the isotropic equivalent kinetic energy E$_{\rm{k,iso}}$ of the outflow; (2) the density of the medium, where either an ISM-like medium ($n_{\rm CSM}$ constant), or a wind-like medium ($\rho_{\rm CSM}=\dot{M}/(4\pi R^{2}\,v_{\rm w})$) produced by a constant progenitor mass-loss rate $\dot M$ can be chosen; (3) the microphysical shock parameters $\epsilon_B$ and $\epsilon_e$, which are the post-shock energy fraction in the magnetic field and electrons respectively \citep[see][for more details]{Sironi15}; (4) the jet opening angle $\theta_{\rm{j}}$, and (5) the observer angle with respect to the jet axis $\theta_{\rm{obs}}$ (hereafter referred to as the ``observer angle''). We fixed the power-law index of the shocked electron energy distribution to $p=2.5$, as it typically varies in the range 2--3 from GRB afterglow modelling (e.g., \citealt{Curran10} and \citealt{Wang15b}). Unless otherwise specified we will report mass-loss rates $\dot M$ for an assumed wind velocity of $v_{\rm w}=1000\,\rm{km\,s^{-1}}$ which is representative of compact massive stars like Wolf-Rayet stars.

We explore two physical scenarios for the interstellar medium, namely ISM-like ($10^{-3}\,{\rm cm}^{-3}\leq n_{\rm CSM}\leq10^2\,{\rm cm}^{-3}$) and wind-like ($10^{-8}\,M_{\odot}\, {\rm yr}^{-1}\leq\dot{M}\leq10^{-3}\,M_{\odot} {\rm yr}^{-1}$), for two jet collimation angles ($\rm{\theta_j=5\arcdeg}$ and $\rm{\theta_j=30\arcdeg}$), three observer angles ($\rm{\theta_{obs}}=30\degr$, 60$\degr$ and 90$\degr$) and isotropic kinetic energies in the range $10^{50}$ erg $\leq E_{\rm k,iso}\leq 10^{55}$ erg.  These values are representative of the parameters that are derived from accurate modeling of the broad-band afterglows of GRBs \citep[e.g.,][]{Schulze11,Laskar13,Perley14,Laskar16}.

In figures \ref{Fig:windconstraints} and \ref{Fig:ismconstraints} we present the results from the entire set of simulations for the range of $\epsilon_e$ and $\epsilon_B$ typically used in the literature. Relativistic shock simulations show $\epsilon_e=0.1$ \citep[e.g.,][]{Sironi15}. $\epsilon_{B}$ is less constrained than $\epsilon_e$. The distribution for $\epsilon_{B}$ derived from GRB afterglow modeling is centered on 0.01 and typically spans $10^{-4}$ to 0.1, with a few claims for smaller values down to $\approx10^{-7}$ (e.g. \citealt{Santana14}). In the text we discuss the results for the fiducial parameters $\epsilon_{\rm e}=0.1$ and $\epsilon_{B}=0.01$, but show the results for other typical values of the micro-physical shock parameters in the figures.

At radio frequencies, the afterglow radiation (i.e., radiation arising from the jet interaction with the medium) consists of synchrotron emission. Both synchrotron emission and synchrotron self-absorption are accounted for in the afterglow models. Free-free absorption is not significant for the CSM densities and blast wave velocities that we consider here. Following \cite{Weiler86}, and considering a wind medium with the highest mass-loss rates investigated here (i.e., $\dot M =10^{-3}\,M_{\odot}\,{\rm yr}^{-1}$) we find the free-free optical depth $\tau_{\rm ff}<0.04$ for frequencies greater than 5 GHz at time $t>26$ days. The SLSN-I 2017egm was observed at 1.6 GHz at $\sim 39$ days since explosion (Table \ref{Tab:Sample}). In this case we estimate a $< 15$\% flux reduction due to free-free absorption for the largest densities considered in this study, with no impact on our major conclusions. For the ISM-like densities considered below, free-free absorption is always negligible.

We consider the radio limits from the entire sample of SLSN-I in this analysis. Note that the constraints that we derive are not driven solely by one SLSN-I. Although the limits for SN 2017egm are significantly deeper than the rest of the sample, we only have early-time coverage for this system. As late-time observations are more constraining for off-axis jets, the other systems in our sample still provide meaningful constraints for off-axis jets (see Fig. \ref{Fig:example}). SN 2017egm exploded in June 2017, so our limits only extend to 47 days (considering only the $\sim$8 GHz observations). Radio observations of this object at later times will place the strongest constraints on off-axis jets in SLSNe-I to date. We will consequently continue radio monitoring of SN 2017egm.

\begin{figure*}
	\centering
	\includegraphics[width=0.75\textwidth]{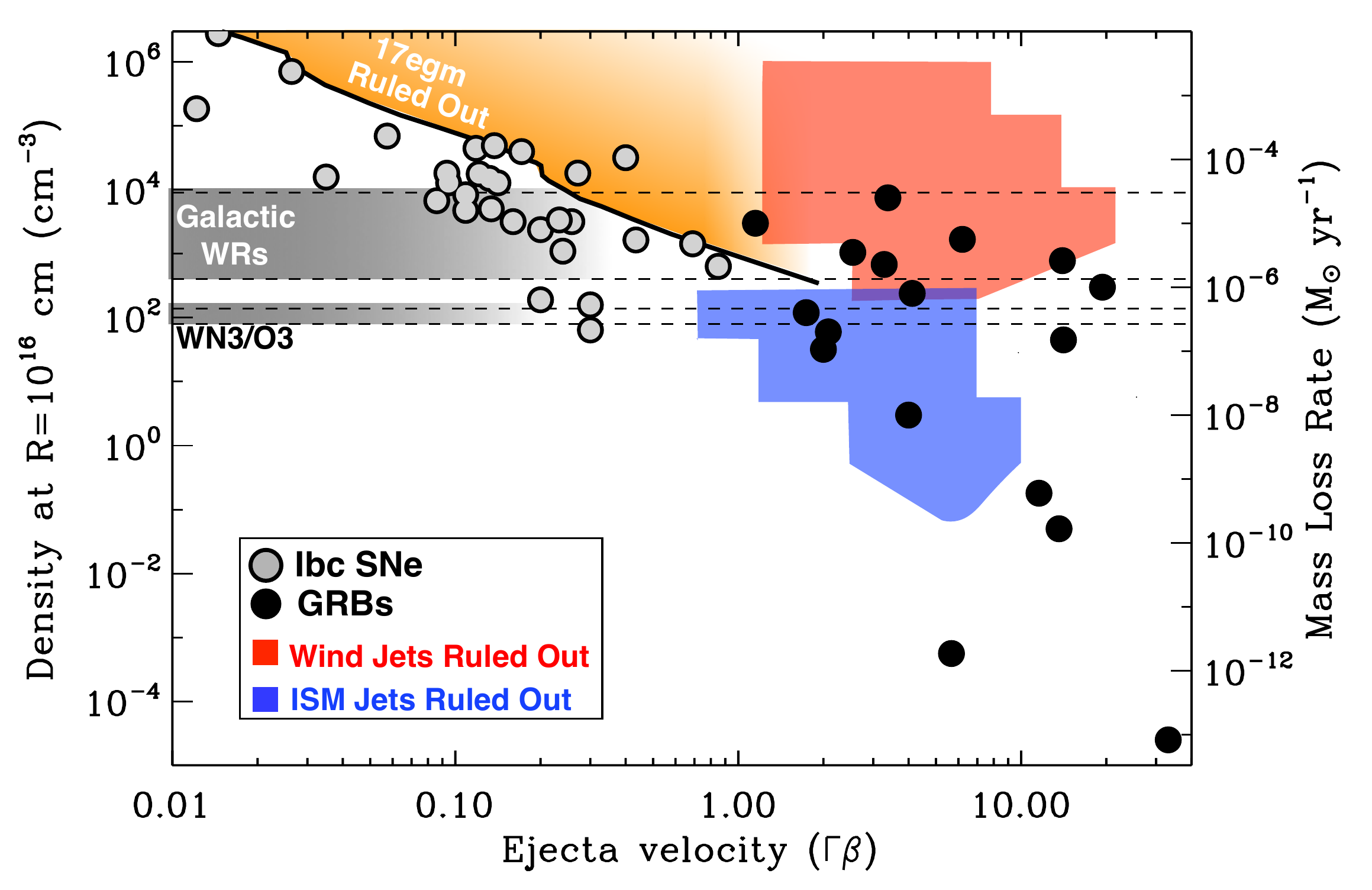}
    \caption{Density in the SN immediate surroundings as a function of the explosion's fastest ejecta for H-stripped core-collapse SN (gray dots) and GRBs (black dots). The red and blue shaded areas mark the region of the parameter space that are ruled out for \emph{any} observer angle by our simulations of relativistic jets (both $\rm \theta_j=30\arcdeg$ and $\rm \theta_j=5\arcdeg$ are included) in wind and ISM environments, respectively (for $\epsilon_e=0.1$, $\epsilon_B=0.01$). Orange-shaded area: region ruled out by the radio limits on SN 2017egm for an uncollimated outflow. The mass-loss scale on the right y-axis is for $v_{\rm w}=1000\,\rm{km\,s^{-1}}$. The velocity of the fast-moving ejecta has been computed at $t=1$ d (rest-frame). Horizontal dashed lines: mass-loss rates measured in the Galaxy for Wolf-Rayet stars \citep{Crowther07} and WN3/O3 stars \citep{Massey15}. References: \citet{Berger03a, Berger03b, Frail06, Soderberg06c, Chandra08, Soderberg08, Cenko10, Soderberg10, Soderberg10b, Cenko11, Troja12, Cano13, Horesh13, Laskar13, Margutti13b,  Margutti14b, Perley14, Walker14, Chakraborti15, Milisavljevic15}.}
    \label{Fig:MassLoss}
\end{figure*}

\subsubsection{Results: $E_{\rm k,iso}$ and $\dot{M}$ phase space}

Figure \ref{Fig:windconstraints} and Figure \ref{Fig:ismconstraints} show the constraints that the upper limits on the radio luminosity of the sample of SLSNe-I place on off-axis jets expanding into a wind profile medium and an ISM profile medium respectively. First consider a wind profile medium, and jets that are off-axis and highly collimated ($\rm{\theta_j=5\arcdeg}$, like those detected in GRBs). For $\epsilon_{\rm e}=0.1$ and $\epsilon_{B}=0.01$ (top-right panel), these off-axis GRB-like jets are ruled out regardless of the observer angle for $\dot{M}\gtrsim10^{-4}\,M_{\odot}\, {\rm yr}^{-1}$ and $E_{\rm k,iso}\gtrsim10^{53}$ erg (this is within the energy range of the observed GRB population). Mass-loss rates such as these are typically found in the winds of extreme red supergiants \citep[e.g.,][]{Smith09a,Smith14} and luminous blue variables \citep[][]{Groh14,Smith14}. To put this in context, Fig.\ \ref{Fig:MassLoss} shows the mass-loss rates and equivalent densities at $10^{16}$ cm (assuming a 1000 km s$^{-1}$ progenitor wind speed) for other H-poor stellar explosions. Specifically, mass-loss rates of this order ($\dot{M}\gtrsim10^{-4}\,M_{\odot}\, {\rm yr}^{-1}$) have been inferred for some SNe-Ibc (Fig.\ \ref{Fig:MassLoss} and references therein), as well as for SN-IIb\footnote{Type IIb SNe are a class that originally show H lines, but transition to type Ib-like (no H lines) over time.} with yellow supergiant progenitors \citep[e.g., SN 2013df\footnote{Note that we assume a 1000 km s$^{-1}$ wind here, and have adjusted the mass loss rate in \citet{Kamble16} accordingly.};][]{Kamble16}. The precluded phase space for highly collimated, off-axis jets with $\dot{M}\gtrsim10^{-4}\,M_{\odot}\, {\rm yr}^{-1}$ and $E_{\rm k,iso}\gtrsim10^{53}$ is indicated in Figure \ref{Fig:MassLoss}. GRB-like jets are not ruled out for the lower-density environments inferred for some GRBs.

If instead, we consider off-axis jets that are less-collimated ($\rm{\theta_ j=30\degr}$) than cosmological GRBs, we can probe to deeper limits as the jet is less collimated to start with and more kinetic energy is coupled to it (with respect to a more collimated jet with same $E_{\rm k,iso}$). In this case, regardless of the observing angle, we can rule out scenarios where $\dot{M}\gtrsim10^{-5}\,M_{\odot} \, {\rm yr}^{-1}$ and $E_{\rm k,iso}\gtrsim10^{53}$ erg ($E_{\rm k}<10^{50}$ erg) as shown in Fig. \ref{Fig:windconstraints} (for $\epsilon_e=0.1$ and $\epsilon_B=0.01$). This parameter space is illustrated in Figure \ref{Fig:MassLoss}. A significant fraction of the galactic Wolf-Rayet population \citep{Vink05,Crowther07} and luminous blue variables \citep[e.g.,][]{Vink02,Smith04}, as well as the most luminous O-type stars \citep[e.g.,][]{deJager88,vanLoon05} show mass loss rates in this range. Off-axis jets of this kind with $E_{\rm k,iso}\gtrsim10^{53}$ erg would also be precluded in the most dense environments inferred for GRBs, and for most of the observed population of hydrogen stripped SNe (assuming a 1000 km s$^{-1}$ wind).

In Section \ref{sec:on-axis_jets} we discussed how this sample of SLSNe-I ruled out on-axis jets of the kind seen in low-luminosity (less-collimated) GRBs with the exception of GRB\,060218. Now consider less-collimated jets ($\rm{\theta_{\rm j}=30\degr}$) that are aligned only slightly off-axis --- within 30\arcdeg\ of our line of sight: jets of this kind are ruled out down to clean environments of $\dot{M} \gtrsim 10^{-8}\,M_{\odot} \, {\rm yr}^{-1}$ where $E_{\rm k,iso}\gtrsim10^{51}$ erg. (Fig.\ \ref{Fig:windconstraints}, for $\epsilon_e=0.1$ and $\epsilon_B=0.01$). Assuming a progenitor wind speed of 1000 km s$^{-1}$, this parameter space precludes the environments of all the detected SN\,Ibc and most of the GRBs detected to date (see Figure \ref{Fig:MassLoss}).

For comparison, Figure \ref{Fig:ismconstraints} gives the equivalent constraints for a constant density environment (modeling of GRB afterglows sometimes indicates a better fit to ISM environments, e.g., \citealt{Laskar14}). For $\epsilon_e=0.1$ and $\epsilon_B=0.01$ (top-right panel), a collimated jet with $\rm{\theta_{\rm j}=5\arcdeg}$ is ruled out regardless of the observer angle for $n_{\rm CSM} \gtrsim 10\,{\rm cm}^{-3}$ and $E_{\rm k,iso}\gtrsim10^{51}$ erg. A jet with $\rm{\theta_{\rm j}=30\arcdeg}$ is ruled out for $n_{\rm CSM } \gtrsim 1 \, {\rm cm}^{-3}$ and $E_{\rm k,iso}\gtrsim10^{51}$ erg. Deeper constraints are obtained for jets with their axes aligned within $30\degr$ or $60\degr$ of our line of sight. Specifically, the jets with $\rm{\theta_{\rm j}=5\degr}$ and observer angles of $\leq30\degr$ are excluded down to $n_{\rm{CSM}} \gtrsim 10^{-3}\,\rm{cm^{-3}}$. 

\begin{figure*}
	\centering
	\includegraphics[width=0.75\textwidth]{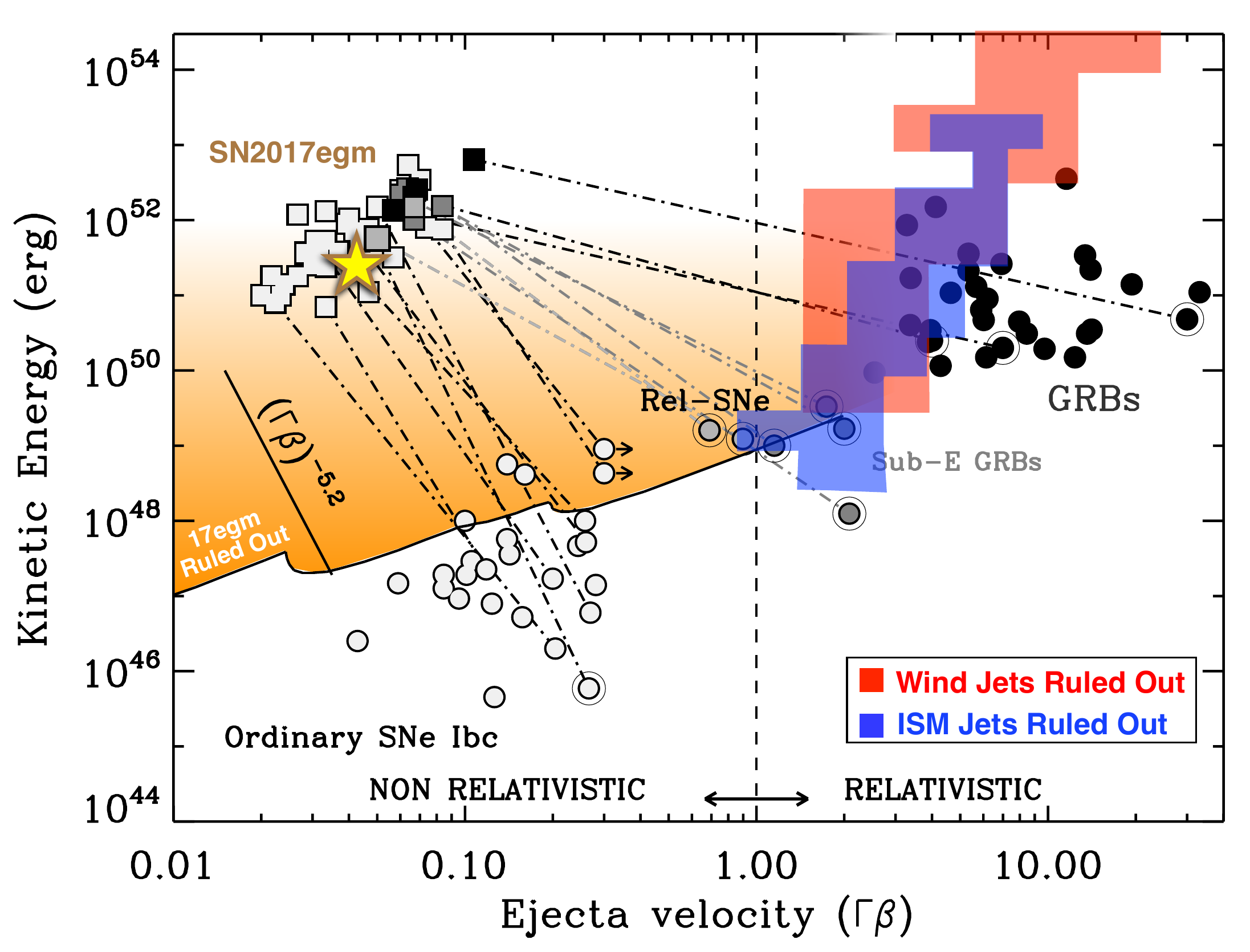}
    \caption{Kinetic energy profile of the ejecta of H-poor cosmic explosions, including ordinary type Ibc SNe, relativistic SNe, GRBs and sub-energetic GRBs. The symbol colors indicate the class of object, namely black for GRBs, gray for relativistic SNe and white for ordinary SNe Ibc. Shaded areas mark the constraints on the properties of SLSNe-I fastest ejecta. Squares and circles are used for the slow-moving and the fast-moving ejecta, respectively, as measured from optical (slow ejecta) and radio (fast ejecta) observations. An additional circle surrounding a point indicates the object showed a broad-lined optical spectrum. The velocity of the fast-moving ejecta has been computed at  $t=1$ d (rest-frame). The ejecta kinetic energy profile of a pure hydrodynamical explosion is also marked as a reference ($E_k \sim (\Gamma\beta)^{-5.2}$, \citealt{Tan01}). The blue and red areas identify the region of the parameter space of the fast moving ejecta that is ruled out based on our simulations of relativistic jets expanding in an ISM and wind-like environments, respectively (for $\epsilon_e=0.1$ and $\epsilon_B=0.01$). Only jet models that are ruled out for \emph{any} observer angle are shown here. Orange shaded area: region of the parameter space that is ruled out based on our simulations of radio emission from non-collimated outflows and the radio limits on SN\,2017egm. The location of the slowly moving ejecta of SN\,2017egm is shown with a star. References: \citet{Berger03a, Berger03b, Frail06, Soderberg06c, Chandra08, Soderberg08, Cenko10, Soderberg10, Soderberg10b, Cenko11, Ben-Ami12, Sanders12, Troja12, Cano13, Horesh13, Laskar13, Margutti13b, Mazzali13, Milisavljevic13, Xu13, Corsi14, Guidorzi14, Kamble14, Margutti14b, Perley14, Walker14, Chakraborti15, Milisavljevic15, Nicholl17b, Bose17}.}
    \label{Fig:EkVelocity}
\end{figure*}

\subsubsection{Results: $E_{\rm k}$ and $\Gamma\beta$ phase space}

Engine-driven explosions (ie., GRBs, sub-energetic GRBs and relativistic SNe) are clearly distinguished from normal spherical core-collapse SNe by a flatter kinetic energy profile of their ejecta \citep[e.g.,][]{Soderberg06c}. For an engine-driven explosion, a larger fraction of the kinetic energy is contained in the fast-moving ejecta than in the slow-moving ejecta---in contrast to a hydrodynamical explosion. This is illustrated in Figure \ref{Fig:EkVelocity}, where we plot the kinetic energy in the slow- and fast-moving ejecta (joined by a dashed line to guide the eye) for the H-poor explosions where these properties have been measured. For a pure hydrodynamical explosion we would expect a profile of $E_{\rm k}\sim (\Gamma\beta)^{-5.2}$, while GRBs have significantly flatter profiles \citep{Tan01}. A flat energy profile for the SLSNe-I would suggest an engine-driven explosion.

We have excluded a region of the $\rm E_k$ versus $\Gamma\beta$ phase-space in Figure \ref{Fig:EkVelocity} for this sample of SLSNe-I based on the limits from our simulations. Specifically, for a collimated ($\rm{\theta_{\rm j}=30\degr}$) jet (ruled out at all observer angles), we deduce limits based on the excluded combinations of mass-loss rate and isotropic kinetic energy (Figs.\ \ref{Fig:windconstraints} and \ref{Fig:ismconstraints}) as follows: applying the standard formulation of the fireball dynamics with expansion in a wind-like and ISM-like environment (e.g., \citealt{Chevalier00}), the bulk Lorentz factor of the downstream fluid behind the shock front is \begin{equation*}
\Gamma\sim 18.7 (E_{\rm{k,iso}}/10^{54}\,\rm{erg})^{1/4} (A_*/0.1)^{-1/4}(t/1\,\rm{day})^{-1/4}
\end{equation*}
for a wind profile medium, and 
\begin{equation*}
\Gamma\sim 10.1 (E_{\rm{k,iso}}/10^{54}\,\rm{erg})^{1/8}(n_{\rm{CSM}}/0.1 \rm{cm^{-3}})^{-1/8}(t/1\,\rm{day})^{-3/8}
\end{equation*}
for an ISM profile medium.
$\Gamma_s=\sqrt[]{2}\Gamma$ is the shock Lorentz factor, $A_*$ is the wind parameter characterizing the density of the wind-generated CSM, and $A_*=1$ for $\dot M=10^{-5}\, M_{\sun} \, {\rm yr}^{-1}$ and $v_{\rm w}=1000\,\rm{km\,s^{-1}}$. In Fig. \ref{Fig:EkVelocity} we plot the beaming corrected kinetic energy $E_{\rm{k}}=E_{\rm{k,iso}}(1-\cos\theta_{\rm j})$ and estimate the specific momentum of the fastest ejecta at an arbitrary time of 1 day post-explosion (rest-frame). The excluded phase space for a jet collimated to $\rm{\theta_{\rm j} \leq 30\degr}$ are shaded in red and blue in Figure \ref{Fig:EkVelocity} for a wind and ISM medium respectively.

The excluded phase space does not constrain the slope of the kinetic energy profile to the extent where we can confirm or rule out the presence in a central engine in this sample of SLSNe-I. Based on our simulations, we ruled out off-axis collimated ($\theta_{\rm j}=30\degr$) jets at $\dot{M}\gtrsim10^{-5}\,M_{\odot}\,{\rm yr}^{-1}$ and $E_{\rm k,iso}\gtrsim10^{52}$ erg for every observing angle. GRB-like jets exploding in less dense environments than we rule out will have faster moving ejecta, and thus appear to the right of the excluded phase space in this figure. This phase space associated with faster moving ejecta is not ruled out because these jets are associated with large $E_{\rm k,iso}$ and very low densities. As the radio emission is produced in the shock front between the jet and the CSM, at low densities the radio luminosity will be lower and more difficult to rule out, especially if it is off-axis.

\section{Constraints on uncollimated outflows}\label{sec:uncollimated_outflows}

Despite the fact that SLSNe-I are significantly more luminous ($\sim10-100$ times) than `normal' type-Ic SNe at optical wavelengths, the deepest SLSN-I limits indicate that they can be significantly fainter than even some normal SNe Ic at radio wavelengths (see Fig. \ref{Fig:SLSNeGRBSNe}). Here we analyze our radio limits in the context of uncollimated (spherical) outflows.

Among H-stripped core-collapse SNe without collimated outflows, relativistic SNe qualify as a separate class. Relativistic SNe are characterized by mildly-relativistic ejecta, bright radio emission---but faint X-ray emission that clearly sets relativistic SNe apart from sub-energetic GRBs---and a kinetic energy profile $E_{\rm k}(\Gamma\beta)$ (Fig.\ \ref{Fig:EkVelocity}) that is suggestive of the presence of a central engine driving the explosion \citep{Soderberg10,Bietenholz10,Chakraborti11,Margutti14b,Chakraborti15}. These observed properties are attributed to a scenario where a jet is present but fails to successfully break through the stellar envelope, possibly due to a shorter-lived engine, or a larger envelope mass  (\citealt{Lazzati12,Margutti14b}, see also \citealt{Mazzali08}). To date only two relativistic SNe, 2009bb and 2012ap, and one candidate (iPTF17cw, \citealt{Corsi17}) are known. As there is no evidence for beaming of the radio emission from relativistic SNe \citep{Soderberg10b,Bietenholz10,Chakraborti15}, the radio limits on SLSNe-I 2017egm, Gaia16apd, and to a lesser extent 2012il, clearly rule out the radio luminosities associated with relativistic SNe (Fig.\ \ref{Fig:SLSNeGRBSNe}). These observations indicate some key difference of the blastwave and environment properties of SLSNe-I and relativistic SNe that we quantify below with simulations of the radio emission from uncollimated ejecta.

The SN shock interaction with the medium, previously sculpted by the stellar progenitor mass loss, is a well known source of radio emission in young SNe \citep[e.g.,][]{Chevalier82,Weiler86,Chevalier06}. The SN shock wave accelerates CSM electrons into a power-law distribution $N(\gamma)\propto \gamma^{-p}$ above a minimum Lorentz factor $\gamma_m$. Radio non-thermal synchrotron emission originates as the relativistic electrons gyrate in amplified magnetic fields. The result is a bell-shaped radio spectrum peaking at frequency $\nu_p$ and cascading down to lower frequency as the medium becomes optically thin to synchrotron self-absorption (SSA). In the case of H-stripped core-collapse SNe, $p\sim 3$ is usually inferred, and SSA dominates over free-free absorption (e.g., \citealt{Chevalier06}). The self-absorbed radio spectrum scales as $F_{\nu} \propto \nu^{5/2}$ below $\nu_p$ and $F_{\nu}\propto \nu^{-(p-1)/2}$ above $\nu_p$ \citep{Rybicki79}. 

We generated a grid of radio spectral models with $p=3$, $\nu_p$ between 0.1-60 GHz and peak spectral luminosity $L_{\nu_p}$ in the range $5\times 10^{26}-10^{29}\,{\rm erg\,s^{-1}Hz^{-1}}$. The comparison to the SLSNe-I radio limits (at all frequencies) leads to robust constraints on the time-averaged velocity of the shock wave ($\Gamma\beta$ in Fig.\ \ref{Fig:SLSNeGRBSNe}), total energy required to power the radio emission ($E$), amplified magnetic field ($B$) and progenitor mass-loss rate ($\dot M$). Our calculations assume a wind-like medium with fiducial miscrophysical parameters $\epsilon_e=0.1$ and $\epsilon_B=0.01$. Following \cite{Chevalier98,Chevalier06,Soderberg12}, for SSA-dominated SNe, the shock-wave radius is given by 
\begin{align*}
R\approx& 3.3\times10^{15}(\epsilon_e/\epsilon_B)^{-1/19}(L_{\nu_p}/10^{26}\,\rm{erg\,s^{-1}Hz^{-1}})^{9/19}\\
&\times(\nu_p/ 5\,\rm{GHz})^{-1}\,\,cm,\\
B\approx& 0.70(\epsilon_e/\epsilon_B)^{-4/19}(L_{\nu_p}/10^{26}\,\rm{erg\,s^{-1}Hz^{-1}})^{-2/19}\\
&\times(\nu_p/ 5\,\rm{GHz})\,\,G,\\
E=&B^2R^3/12\epsilon_B,\\
\dot M\approx& 0.39\times 10^{-5}(\epsilon_B/0.1)^{-1} (\epsilon_e/\epsilon_B)^{-8/19}\\
&\times(L_{\nu_p}/10^{26}\,\rm{erg\,s^{-1}Hz^{-1}})^{-4/19}\\
&\times(\nu_p/ 5\,\rm{GHz})^{2}\\
&\times(t_{p}/10\,\rm{days})^2\,\,\rm{M_{\sun}yr^{-1}}.\\
\end{align*} 

Based on these simulations we find that the radio limits on the SLSN-I 2017egm produce interesting constraints in the  $\dot M$, $E_{\rm{k}}$ and $\Gamma\beta$ phase space (orange-shaded area in Figs.\ \ref{Fig:MassLoss} and \ref{Fig:EkVelocity}). At any given velocity of the fastest ejecta, the limits on SN 2017egm rule out $E_{\rm k}>10^{48}$ erg coupled to the fastest ejecta (Fig.\ \ref{Fig:EkVelocity}), and the densest environments found in association with H-stripped core-collapse SNe (Fig. \ref{Fig:MassLoss}). Current limits however do not constrain the slope of the $E_{\rm{k}}(\Gamma\beta)$ profile and do not rule out the region of the parameter space populated by spherical hydrodynamical collapses with $E_{\rm{k}}<10^{48}$ erg in their fastest ejecta (Figure \ref{Fig:EkVelocity}).

\section{Summary and Conclusions}\label{sec:Conclusions}

We have compiled all the radio observations of SLSNe-I published to date and presented three new observations (a sample of nine SLSN-I). Based on these limits, we constrain the sub-pc environments and fastest ejecta in this sample of SLSN-I for the case that a relativistic jet or an uncollimated outflow were present. For this analysis we make the necessary assumption that the jet/environment properties do not vary within this sample of SLSN-I. These are our main results:

\begin{itemize}
\item In this sample of SLSNe-I we rule out collimated on-axis jets of the kind detected in GRBs.\\
\item We do not rule out the entire parameter space for off-axis jets in this sample, but do constrain the energies and circumstellar environment densities if off-axis jets were present.\\
\item If the SLSNe-I in this sample have off-axis GRB-like (collimated to $\rm{\theta_j=5\arcdeg}$) jets, then the local environment is of similar (or lower) density to that of the detected GRBs. Specifically, if off-axis jets of this kind were present, then they had energies $E_{\rm k,iso}<10^{53}$ erg ($E_{\rm k} <4\times10^{50}$ erg) in environments shaped by progenitors with mass loss rates $\dot{M}<10^{-4}\,M_{\odot}\, {\rm yr}^{-1}$ (for microphysical shock parameters $\epsilon_e=0.1$ and $\epsilon_B=0.01$). This would, for example, exclude jets with $E_{\rm k} >4\times10^{50}$ erg if the progenitor mass-loss rates were of the order typically found in the winds of extreme red supergiants and luminous blue variables (and inferred for some SNe-Ibc and SNe-IIb).\\
\item If this sample of SLSNe-I produced off-axis jets that are less collimated ($\rm{\theta_j=30\degr}$) than cosmological GRBs, then the jets must have energies $\rm{E_{k,iso}<10^{53}}$ erg ($E_{\rm k}<10^{50}$ erg) and occur in environments shaped by progenitors with mass loss rates $\dot{M}<10^{-5}\,M_{\odot}\,{\rm yr}^{-1}$. This precludes jets of this kind with $E_{\rm k}<10^{50}$ erg for the mass-loss rates inferred for most of the observed population of hydrogen stripped supernovae, and the most dense environments inferred for GRBs.\\
\item The deepest SLSNe-I limits rule out emission from faint un-collimated GRBs (`sub-luminous' or `sub-energetic' GRBs) including GRB\,980425, but with the exception of GRB\,060218. To successfully probe emission at the level of all poorly-collimated GRBs (like GRB\,060218), radio observations of SLSNe-I in the local Universe ($z\leq0.1$) need to be taken $\leq10$ days after explosion.\\
\item The radio limits of this sample of SLSNe-I rule out radio luminosities associated with relativistic SNe, and thus likely indicate some key difference of the blast wave and environment properties between these two classes of objects. Significant differences in the respective structures of the progenitor stars and/or jet longevity may also affect the ability of the central engine's jet to successfully break through the stellar envelope.\\
\item We only partially rule out the phase space associated with uncollimated outflows such as the kind seen in ordinary type Ibc SNe.\\
\item For SN 2017egm (the closest SLSN-I observed at radio wavelengths to date) we can constrain the energy of a possible uncollimated outflow to $\rm{E_k\lesssim10^{48}\,erg}$, which is consistent with the kinetic energy associated with the fastest moving material in ordinary type Ibc SNe.\\
\end{itemize}

Radio observations of SLSNe-I are a powerful tool to constrain the central engine properties and sub-parsec environments of these luminous, H-stripped stellar explosions. To fully constrain the outflow properties and environment properties, a combination of both early time and late time coverage is needed. Specifically, to probe emission at the level of all poorly-collimated GRBs (like GRB\,060218), radio observations of SLSNe-I in the local Universe ($z\leq0.1$) need to be taken $\leq10$ days after explosion. Late time radio observations taken hundreds of days post-explosion are necessary to detect off-axis collimated jets in these systems. We will be able to constrain this much more efficiently with more sensitive radio telescopes that are coming online or planned, such as MeerKAT, the next generation VLA (ngVLA, \citealt{ngVLA}) and the Square Kilometer Array (SKA, \citealt{SKA}).

\bigskip
This research made use of the Open Supernova Catalog \citep{Guillochon17}, and NASA's Astrophysics Data System Bibliographic Services. The National Radio Astronomy Observatory is a facility of the National Science Foundation operated under cooperative agreement by Associated Universities, Inc. C.G. acknowledges University of Ferrara for use of the local HPC facility co-funded by the ``Large-Scale Facilities 2010'' project (grant 7746/2011). We thank University of Ferrara and INFN-Ferrara for the access to the COKA GPU cluster. Development of the Boxfit code was supported in part by NASA through grant NNX10AF62G issued through the Astrophysics Theory Program and by the NSF through grant AST-1009863. Simulations for BOXFIT version 2 have been carried out in part on the computing facilities of the Computational Center for Particle and Astrophysics (C2PAP) of the research cooperation "Excellence Cluster Universe" in Garching, Germany. D. C. and R. M. acknowledge partial support from programs No. 
NNX16AT51G and NNX16AT81G provided by NASA through Swift Guest Investigator Programs. Support for BAZ is based in part while serving at the NSF. At the time of the observations LC was a Jansky Fellow of the National Radio Astronomy Observatory. G. M. acknowledges funding support from the UnivEarthS Labex program of
Sorbonne Paris Cit\'e (ANR-10-LABX-0023 and ANR-11-IDEX-0005-02) and from the French Research National Agency: CHAOS project ANR-12-BS05-0009.

\bibliographystyle{apj}
\bibliography{sne}

\appendix
\section{Brief description of the SLSNe-I in our sample}\label{appendix:descriptions}

\subsection{PTF09cnd}

PTF09cnd was detected by the Palomar Transient Factory during commissioning, and was one of the systems that originally defined the SLSN class \citep{Quimby11}. Properties of the host galaxy are given in \citet{Neill11} and \citet{Perley16}. X-ray observations at 0.3-10 keV produced non-detections with unabsorbed fluxes in the range $10^{-15}-10^{-13}$ erg\,s$^{-1}$\,cm$^{-2}$ \citep{Levan13,Margutti17b}. Radio observations of PTF09cnd are reported on in \citet{Chandra09ATel} and \citet{Chandra10ATel} (see Table \ref{Tab:Sample}).

\subsection{PS1-10awh}

PS1-10awh was discovered in October 2010 in the Pan-STARRS1 Medium Deep Survey, while on the rise to peak \citep{Chomiuk11}. UV/optical photometry and time-series spectroscopy from day $-21$ to 26 is presented in \citet{Chomiuk11}. \citet{Lunnan14} detected the host galaxy at $M_{\rm{F606W}}=27$ mag with the Hubble Space Telescope. \citet{Chomiuk11} observed PS1-10awh at radio wavelengths (see Table \ref{Tab:Sample}).

\subsection{PS1-10bzj}
PS1-10bzj was discovered in the Pan-STARRS \citep{Chambers16} Medium Deep Survey and classified as a SLSN-I with a peak magnitude of around $-21.4$ mag \citep{Lunnan13}. \citeauthor{Lunnan13} find that the luminosity cannot be powered solely by radioactive Nickel decay, but a magnetar or interaction model can explain the observed properties (although intermediate-width spectral lines are predicted by the interaction model, none were detected). Radio observations of this SN are presented here for the first time (see Table \ref{Tab:Observinglog}).

\subsection{PS1-10ky}

PS1-10ky was discovered near peak brightness in the Pan-STARRS1 Medium Deep Survey, at $z=0.9558$ \citep{Chomiuk11}. At peak, its bolometric magnitude was $-22.5$ mag (see \citealt{Chomiuk11}). No host has been detected for this SN \citep{Chomiuk11,Lunnan15,McCrum15}. Radio observations of PS1-10ky have been presented in \citet{Chomiuk11} -- see Table \ref{Tab:Sample}.

\subsection{SN 2012il}

SN 2012il (alternatively PS1-12fo or CSS120121:094613+195028) was independently discovered in the Pan-STARRS1 3Pi Faint Galaxy Supernova Survey \citep{Smartt12ATel} and the Catalina Real-time Transient Survey \citep{Drake12ATel} on 2012 January 19 and 21 respectively. \citet{Smartt12ATel} classified it as a SLSN-I at $z=0.175$. Based on the late-time luminosity decline rate, and the short diffusion times at peak, \citet{Inserra13} favor the magnetar model. See \citet{Lunnan14} for optical spectroscopy of the host galaxy. X-ray observations of SN 2012il in the 0.2--10 keV band yielded non-detections with upper-limits on the unabsorbed flux in the range $10^{-13}-10^{-12}$ erg cm$^-2$ s$^{-1}$ \citep{Levan13,Margutti17}. The radio observations of this system were taken by \citet{Chomiuk12} -- see Table \ref{Tab:Sample}.

\subsection{SN 2015bn}

SN 2015bn (alternatively PS15ae, CSS141223:113342+004332 or MLS150211:113342+004333) was discovered by the Catalina Real-time Transient Survey \citep{Drake09} and the Pan-STARRS Survey for Transients \citep{Huber15ATel}. It was classified as a SLSN-I by \citet{Guillou15ATel}, which was confirmed by \citet{Drake15ATel}.
In UV observations (obtained with \textit{Swift}) to NIR observations from $-50$ to 250 days from optical peak, \citet{Nicholl16} found it was slowly evolving and showed large undulations (on timescales of $30-50$ days) superimposed on this evolution in the light curve. The nature of these undulations is discussed in \citet{Nicholl16} and \citet{Yu17}.

Nebular phase observations of SN 2015bn \citep{Nicholl16b,Jerkstrand17} indicate that it is powered by a central engine \citep{Nicholl16b}. From spectropolarimetric observations of the SN, \citet{Inserra16} inferred an axisymmetric geometry and found that the evolution could be consistent with a central engine. Additional polarimetric observations can be found in \citet{Leloudas17}. X-ray observations at 0.3-10 keV produced non-detections with unabsorbed fluxes that were predominantly in the range $10^{-14}-10^{-13}$ erg\,s$^{-1}\,cm^{-2}$ \citep{Nicholl16,Inserra17,Margutti17b}. Radio observations of SN 2015bn are presented in \citet{Nicholl16} (see Table \ref{Tab:Sample}).

\subsection{iPTF15cyk}
iPTF15cyk was detected on 2015 September 17 in optical follow-up observations of the gravitational wave source GW150914 by the intermediate Palomar Transient Factory \citep{Kasliwal16}. Further spectroscopic and multi-wavelength observations confirmed it as a SLSN-I \citep{Kasliwal16}. X-ray observations \citep{Kasliwal16} gave an upper-limit on the unabsorbed flux of $1.3\times10^{-13}$ erg cm$^{-2}$ s$^{-1}$. Radio observations (\citet{Palliyaguru16} and \citet{Kasliwal16}, see Table \ref{Tab:Sample}) were taken as part of the first broadband campaign to search for an electromagnetic counterpart to an Advanced Laser Interferometer Gravitational-Wave Observatory (LIGO) gravitational wave trigger \citep{Abbott16}.

\subsection{Gaia16apd}

Gaia16apd (alternatively SN 2016eay) was discovered by the Gaia Photometric Survey (Gaia collaboration 2016\nocite{GAIA-collaboration16}) and classified as a SLSN-I by \citet{Kangas16}. At early times it was extremely bright at UV wavelengths (peaking at $−23.3$ mag) in comparison to other SN classes \citep{Blagorodnova16,Kangas17,Yan17}. Based on later time UV (observations obtained with \textit{Swift}) and optical observations, \citet{Nicholl17} concluded that this is due to a more powerful central engine rather than interaction with the surrounding medium or a lack of UV absorption. \citeauthor{Nicholl17} found that the best fit for a magnetar central engine has a spin period of 2 ms and magnetic field of $4\times10^{14}$ G (in agreement with \citealt{Kangas17}). Radio observations of Gaia16apd are presented here for the first time (see Table \ref{Tab:Observinglog}).

\subsection{SN 2017egm}

SN 2017egm (alternatively Gaia17biu) was discovered by the Gaia satellite on 2017 May 26, and was originally classified as a Type II SN \citep{Xiang17ATel}. \citet{Dong17ATel} then re-classified it as a SLSN-I at $z=0.030721$, making it the closest SLSN-I discovered to date.

Although an initial X-ray detection of the SN was claimed \citep{Grupe17ATel}, it was later found that the X-ray emission was more likely associated with star formation in the host galaxy NGC 3191 \citep{Grupe17ATelb,Bose17}. Several radio observations were taken at different frequencies (see Table \ref{Tab:Sample} for details), but SN 2017egm was not detected \citep{Romero-Canizales17ATel,Bright17ATel,Coppejans17ATel,Bose17}.

The host galaxy, NGC 3191, is unlike those of other SLSN-I\@. It is a metal rich ($\approx1.3\rm{Z}_{\odot}$ at the radial offset of the SN) massive spiral galaxy, with a high star formation rate of $\approx 15 \, M_{\odot}\,\rm{yr}^{-1}$ and mass of $\approx 10^{10.7} M_{\odot}$ (\citealt{Nicholl17b}, see also \citealt{Bose17}). \citet{Izzo17} however, show that there are spatial variations in the host properties, and the local environment of SN\,2017egm is not so unusual. NGC 3191 is also a known radio source: It was detected at 1.4 GHz in the FIRST survey (Faint Images of the Radio Sky at Twenty-Centimeters) with a peak flux density of $1.63\pm0.13$ mJy beam$^{-1}$ and an integrated flux density of $15.96\pm0.13$ mJy \citep{White97}. Furthermore, NGC 3191 was detected at $1.8\pm0.1$ mJy at 15.5 GHz \citep{Bright17ATel}, and was detected and resolved at 10 GHz \citep{Bose17}. In our 33 GHz observations we do not detect the host galaxy to a 3-sigma upper-limit of 150 $\mu$Jy beam$^{-1}$.
\end{document}